\documentclass[twocolumn,aps,pra]{revtex4}
\usepackage{epsfig}
\usepackage[english]{babel}
\usepackage{latexsym}
\usepackage{graphics}
\usepackage{subfigure}
\usepackage{epsfig}
\usepackage{graphicx}
\usepackage{dcolumn}
\usepackage{amsmath}
\usepackage{hyperref}
\usepackage{color}

\begin{document}

\title{Strong-field double ionization dynamics of vibrating HeH$^+$ versus HeT$^+$}
\author{S. Wang$^{1}$, R. H. Xu$^{2}$, W. Y. Li$^{3,*}$, X. Liu$^{4}$, W. Li$^{4}$,  G. G. Xin$^{5,\dag}$,  Y. J. Chen$^{1,\ddag}$}
\affiliation{1.College of Physics and Information Technology, Shaan'xi Normal University,   Xi'an, China\\
2.Institute of Applied Physics and Computational Mathematics, Beijing 100091, China\\
3.School of Mathematics and Science, Hebei GEO University, Shijiazhuang 050031, China\\
4.Beijing Institute of Space Mechanics and Electricity, Chinese Academy of Space Technology, Beijing, 100094, China\\
5.School of Physics, Northwest University, Xi'an 710127, China}

\date{\today}

\begin{abstract}
We study double ionization (DI) dynamics of vibrating HeH$^+$ versus its isotopic variant HeT$^+$ in strong laser fields numerically.
Our simulations show that for both cases, these two electrons in  DI  prefer to release together along the H(T) side.
At the same time, however, the single ionization (SI) is preferred when the first electron escapes along the He side.
This potential mechanism is attributed to
the interplay of the rescattering of the first electron and the Coulomb induced large ionization time lag.
On the other hand, the nuclear motion increases the contributions of these two electrons releasing together along the He side.
This effect differentiates DI of HeH$^+$ from  HeT$^+$.

\end{abstract}

\maketitle

\section{Introduction}
Strong-laser-matter interaction leads to  many interesting physical processes,
such as above-threshold ionization (ATI) \cite{Yang1993,Lewenstein1995}, high-harmonic generation (HHG) \cite{McPherson1987,Ferray1988,Lewenstein1994},  double and multiple ionization \cite{Huillier1983,Walker1994,Palaniyappan2005,Becker2012,Corkum2005,Wang2015}, and laser induced electron diffraction \cite{Blaga2002,Wolter2016}, etc., which have promising applications in attosecond science  \cite{Krausz2009}.

Present studies have revealed the importance of tunneling and rescattering in strong-field processes \cite{Schafer1,Corkum1993}.
In comparison with other strong-field processes, double and multiple ionization which involves electron-electron correlation
includes richer physical phenomena.
As the double ionization (DI) from atoms and symmetric molecules have been studied widely \cite{Becker2012},
DI from polar molecules with a large permanent dipole \cite{Etches2010,Dimitrovski2011} is less studied.
Especially, when the nuclear motion is considered, the situation is more complex.
It has been shown that for small polar molecules such as HeH$^+$, the interaction of the strong laser field and the permanent dipole
induces the rapid nuclear motion which has important influences on HHG and ATI of the asymmetric system \cite{wyli2016,Wang2017,Wustelt2018,Yue2018,Li2019}.
In addition, the interplay of the Coulomb effect and the permanent-dipole effect also gives rise to
a strong asymmetry in photoelectron momentum distributions (PMD) for single ionization (SI) of HeH$^+$ \cite{Wang2019}.
This asymmetry is closely related to
the permanent dipole induced asymmetric ionization \cite{Kamta2005} and the Coulomb induced large ionization time delay \cite{Chen2019}.
Effects of these mechanisms on DI of the asymmetric system are not clear so far.

In this paper, we focus on DI  of  $\mathrm{HeH}^{+}$ and its isotopic variant  $\mathrm{HeT}^{+}$
in strong linearly polarized laser fields beyond the Born-Oppenheimer (BO) approximation.
The $\mathrm{HeH}^{+}$ system, the simplest polar heteronuclear molecule,
has served theoretically  and experimentally as a fundamental benchmark system
for understanding molecular formation and electron correlation \cite{Banyard1970}.
Numerical solution of  the time-dependent Schr\"{o}dinger equation (TDSE) of the vibrating two-electron system in full dimensions
is still not within reach and is easily limited by the existing computing capability.
Thus, we use
a simplified model where the motion of all particles is restricted to one dimension (1D).
It has been shown that such a  model can reproduce all important strong-field effects such
as multiphoton ionization and HHG \cite{Eberly1989,Su1991,Schwengelbeck1994}.
This simplified model can  also describe qualitatively  the correlation effects between
electrons and interplay between the electronic and the nuclear motion \cite{Lein2000,Bauer1997,Lappas1998,Lein2002}.

The calculated  PMDs of DI for HeH$^+$ or HeT$^+$ show a striking asymmetry with indicating
that these two electrons in the DI process prefer to release together along the H(T) side.
This phenomenon in DI differs remarkably from that in SI for HeH$^+$ or HeT$^+$,
which shows that the first electron in SI prefers to escape along the He side.
This disagreement between DI and SI strongly implies that the rescattering of the first electron
plays an important role in DI of the asymmetric system.
This rescattering along with the Coulomb induced large ionization time delay remarkably increases
the DI yields and results in preferred DI along the H(T) side.
On the other hand, the rapid nuclear motion,
which differs remarkably for HeH$^+$ and HeT$^+$,
increases the contributions of direct ionization (which is preferred along the He side) to DI.

\section{Numerical Methods}
The Hamiltonian of the asymmetric molecule $\mathrm{HeH}^{+}$ studied here has the following form (in atomic untis of $\hbar=e=m_{e}=1$):
\begin{eqnarray}
\hat{H}(t)=-\frac{\partial R^{2}}{2\mu_{N}}+\frac{Z_{1}Z_{2}}{R}+\frac{1}{\sqrt{(x_{1}-x_{2})^2+\epsilon}}\\\nonumber
+\sum_{j=1}^{2}\Bigg[-\frac{\nabla^2_{x_{j}}}{2\mu_{e}}+V_{en}(R,x_{j})+x_{j}E(t)\Bigg].
\end{eqnarray}
Here R is the internuclear separation and $x_{j}$
(j=1,2) is the electronic coordinate.
 $\mu_{N}=M_{H_{e}}M_{H}/(M_{H_{e}}+M_{H})$ is the nuclear reduced mass and $\mu_{e}=(M_{H_{e}}+M_{H})/(M_{H_{e}}+M_{H}+1)\approx 1$ is the electronic reduced mass. $M_{H_{e}}$ and $M_{H}$ are masses of He and H nuclei.
The term  $V_{en}$ denotes the interaction between the electron and nuclei and has the following form:
\begin{eqnarray}
V_{en}(R,x)=-\frac{Z_{1}}{\sqrt{(x-R_{1})^2+\epsilon}}-\frac{Z_{2}}{\sqrt{(x-R_{2})^2+\epsilon}},
\end{eqnarray}
where $Z_{1}=2$ and $Z_{2}=1$ are the charges for He and H centers, respectively. $R_{1}$ and $R_{2}$ are positions of He and H nuclei with $R_1=M_{H}R/(M_{He}+M_{H})$ and $R_2=-M_{He}R/(M_{He}+M_{H})$. $\epsilon=0.59$ is the smoothing parameter,
which is adjusted such that the ground-state energy of the model HeH$^+$ molecule matches the real one of $E_0=-2.98$ a.u..
The equilibrium separation of  model HeH$^+$  studied here is $R_e=2$ a.u., which also holds for model HeT$^+$
and is somewhat larger than the real one of $R_e=1.4$ a.u..
Here, we have used the length-gauge form of the interaction
Hamiltonian. The laser field used here is $E(t)=E_{0}f(t)\sin(\omega_0 t)$ with peak amplitude $E_{0}$, envelope function f(t)  and laser frequency $\omega_0$. In our simulations, we use a seven-cycle laser pulse which is linearly turned on and off for two optical cycles, and kept at a constant intensity for three additional cycles.

We use $\Psi(R,x_{1},x_{2},t)\equiv\Psi(t)$ on a three-dimensional grid to represent the wave function.
The TDSE of $i\dot{\Psi}(t)=H(t)\Psi(t)$ is solved  numerically using the spectral method \cite{Feit1982}.
A grid size of $L_{x_1}\times L_{x_2}=204.8\times204.8$ a.u. with the grid step of $\Delta x_1=\Delta x_2=0.4$ a.u. for the electron,
a range of R = 0.6...6.9 a.u. with the grid step of $\Delta R=0.1$ a.u. for the internuclear distance, and a time step of $\Delta t=0.05$ a.u.
have proven sufficient convergence for describing the strong field dynamics.
In order to avoid the reflection of the electron wave packet from the boundary
and obtain the momentum space wave function, the coordinate
space is split into the inner and the outer regions with ${\Psi}(t)={\Psi}_{in}(t)+{\Psi}_{out}(t)$,
by multiplication using a mask function  $F(x_1,x_2,R)=F_1(x_1)F_2(x_2)F_3(R)$.
Here,
$F_1(x_1)=\cos^{1/2}[\pi(|{x_1}|-r_0)/(L_{x_1}-2r_0)]$ for $|{x_1}|\geq r_0$ and $F_1(x_1)=1$   for $|{x_1}|< r_0$.
$r_0=3/8L_{x_1}$  is the critical boundary between the inner and the outer regions for one electron.
The relevant electronic wave packet passing through this critical boundary will be absorbed by the mask function smoothly.
The form of $F_2(x_2)$ is similar to $F_1(x_1)$. 
A similar absorbing procedure with the mask function $F_3(R)$ is also used for the upper boundary of $R$.
In the inner region, the wave function ${\Psi}_{in}(t)$ is propagated with the complete Hamiltonian $H(t)$. In the outer region, the time evolution of the wave function ${\Psi}_{out}(t)$ is carried out in momentum space with the Hamiltonian of the free electron in the laser field \cite{Lein2002-2,Tong2006,Henkel2011}. The mask function is applied at each time  interval  of 1 a.u.
and the obtained new fractions of the outer wave function at the DI condition of $|x_1|\geq r_b$ and $|x_2|\geq r_b$,
denoted with  $\Psi_{out}^d(t)$,
are added coherently or non-coherently to the corresponding momentum-space wave function $\tilde{{\Psi}}^{d}_{out}(t)$.
Finally, we obtain PMDs $c_{d}(p_1,p_2)$ of DI from  $\tilde{{\Psi}}^{d}_{out}(t)$. Similarly, with coherently or non-coherently adding the obtained new fractions of the outer wave function at the SI condition of $|x_1|<r_b$  or $|x_2|<r_b$, denoted with $\Psi_{out}^s(t)$, to the corresponding momentum-space wave function $\tilde{{\Psi}}^s_{out}(t)$,
one can obtain  PMDs $c_{s}(p_{1(2)})$ of SI from  $\tilde{{\Psi}}^{s}_{out}(t)$.
Here, $r_b=8$ a.u., which defines the spacial region where the electron is considered to be located at  bound states \cite{Lein2002}.
Accordingly, the integral  of the loss at the DI (SI) grid
boundaries over time gives the total DI (SI) probability P$_{d(s)}$  with
$P_{d(s)}={\int[1-|\Psi_{in}(R,x_{1},x_{2},t)|^{2}]dRdx_{1}dx_{2}dt}={\int\gamma_{d(s)}(R)dR}$ at $|x_1|\geq r_b$ and $|x_2|\geq r_b$
($|x_1|<r_b$  or $|x_2|<r_b$),
which includes contributions $\gamma_{d(s)}(R)$  at different $R$.
Here, we focus on the main characteristics of PMD from HeH$^+$ versus HeT$^+$. For clarity, we present the non-coherent results, i.e.,
$c_{d}(p_1,p_2)={\int|\tilde{{\Psi}}^{d}_{out}(t)|^{2}dRdt}=\int\beta_{d}(R)dR$,
and $c_{s}(p_{1(2)})={\int|\tilde{{\Psi}}^{s}_{out}(t)|^{2}dRdt}=\int\beta_{s}(R)dR$,
which also include R-dependent contributions $\beta_{d(s)}(R)$.
This TDSE treatment for HeT$^+$ is similar to that for HeH$^+$ with replacing H by T in relevant expressions.

\section{Results and discussions}
\subsection{Asymmetric PMDs of DI}
\begin{figure}[t]
\begin{center}
\rotatebox{0}{\resizebox *{8.5cm}{8cm} {\includegraphics {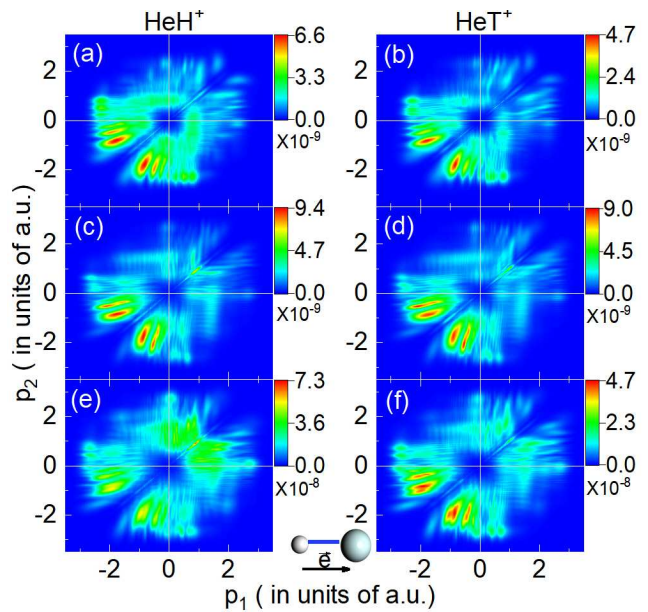}}}
\end{center}
\caption{PMDs of DI for {HeH}$^{+}$  (the left column)  and {HeT}$^{+}$ (right).  The laser parameters used are $I=1\times10^{15}$W/cm$^{2}$, $\lambda=500$ nm for the top row, $I=1\times10^{15}$W/cm$^{2}$, $\lambda=600$ nm for the middle row and $I=1.5\times10^{15}$W/cm$^{2}$, $\lambda=500$ nm for the bottom row. The insets show the positions of these two nuclei with H(T) at the left and He at the right and the unit vector $\vec{e}$ along the laser polarization.}
\label{fig.1}
\end{figure}
In Fig. \ref{fig.1}, we show calculated  PMDs of DI  for $\mathrm{HeH}^{+}$
and  $\mathrm{HeT}^{+}$  at different laser parameters.
These distributions indicate momentum correlation between these two electrons in DI.
Firstly, the distributions in Fig. 1 show a strong asymmetry with the amplitudes in the third quadrant
remarkably larger than those in the first quadrant.
Secondly, when increasing  laser intensities or wavelengthes, the contributions of the first quadrant increase and
this asymmetry becomes smaller.
Thirdly, for the same laser parameters, this asymmetry is smaller for HeH$^+$ than for HeT$^+$.
By contrast,  the distributions in the second and the fourth quadrants are similar for both isotopic cases.
In the following, we concentrate on the origin of this asymmetry,
associated with PMDs of DI in the first and the third quadrants.

\begin{figure}[t]
\begin{center}
\rotatebox{0}{\resizebox *{8.5cm}{8cm} {\includegraphics {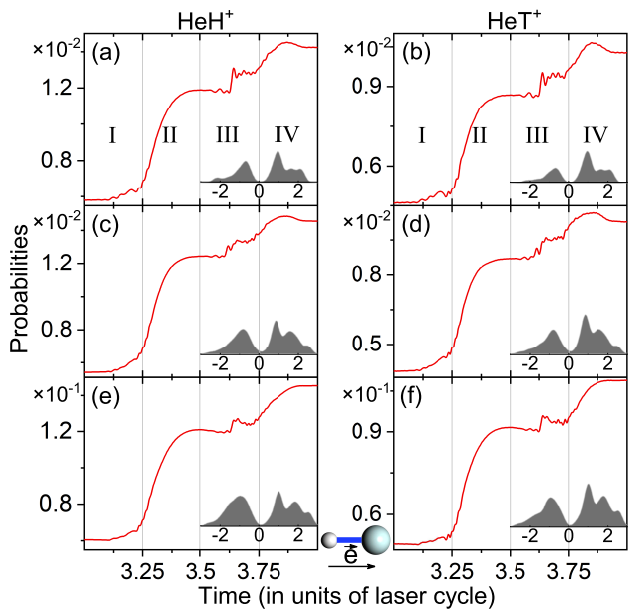}}}
\end{center}
\caption{Time-dependent ionization probabilities of P(t) (red curves) for {HeH}$^{+}$  (the left column)
 and {HeT}$^{+}$ (right) in one laser cycle. In each panel, the inset (shaded area) shows the PMD of SI. The  vertical lines
divide the one-cycle time region into four parts (I-IV). The laser parameters used in each panel are as those in
the corresponding panel in Fig. 1.
} \label{fig.2}
\end{figure}

\subsection{Mechanisms of SI}

To explore the potential mechanism, in Fig. \ref{fig.2}, we plot the time-dependent ionization probabilities in one laser cycle,
which is approximately evaluated with $P(t)=1-\sum_{n=1}^{n=15}|\langle n|\Psi(t)\rangle|^2$ \cite{Chen2012} and is relating to SI.
Here, $|n\rangle$ is the $n$th electronic bound eigenstate of the field-free Hamiltonian $H_0=\frac{1}{\sqrt{(x_{1}-x_{2})^2+\epsilon}}
+\sum_{j=1}^{2}[-\frac{\nabla^2_{x_{j}}}{2\mu_{e}}+V_{en}(R,x_{j})]$ at the BO approximation.
Excluding more bound-state components from $|\Psi(t)\rangle$, results are similar to $P(t)$.
We divide the one-cycle time region into four parts denoted with I-IV.
First, for all cases in Fig. \ref{fig.2}, the ionization  is strong in the first half laser cycle and is weak in the second half laser cycle.
This phenomenon has been termed as asymmetric ionization and is identified as arising from the effect of the permanent dipole \cite{Kamta2005}.
Secondly, in the first half laser cycle, the ionization mainly occurs in the region II after the laser field arrives at its peak.
The reason has been attributed to the Coulomb induced large ionization time delay.
Due to this delay, many electrons which tunnel out of the laser-Coulomb formed barrier near the peak of the laser field
in region I are ionized finally in region II \cite{Chen2019}.
By comparison, in the second half laser cycle, the contributions in region IV after the time of  peak intensity also dominate the ionization for
the same reason of Coulomb induced delay as in the first half cycle, but the contributions of region III are also non-negligible.
In Ref. \cite{Chen2012}, it has been shown that the contributions in region III arise from effects of  excited  states.
Specifically, some electrons are pumped into the excited states from the ground state around the peak of the laser field
in the first half laser cycle and survive the falling part of the laser field of region II.
Then the excited-state electrons with lower ionization potentials are ionized mostly in the arising part of the laser field of region III.
Thirdly, as increasing the laser intensity and wavelength, the contributions of region III decrease due to
the decrease of the excited state effect, as discussed in \cite{Wang2017}.
Fourthly, when the ionization yields of HeH$^+$ with lighter nuclei are larger than  HeT$^+$, a careful analysis tells that the ionization asymmetry in the first and the second half laser cycle
is somewhat more remarkable for HeT$^+$ than for HeH$^{+}$.
These SI characteristics will be used to analyze the potential DI mechanisms.

\begin{figure}[t]
\begin{center}
\rotatebox{0}{\resizebox *{8.5cm}{6cm} {\includegraphics {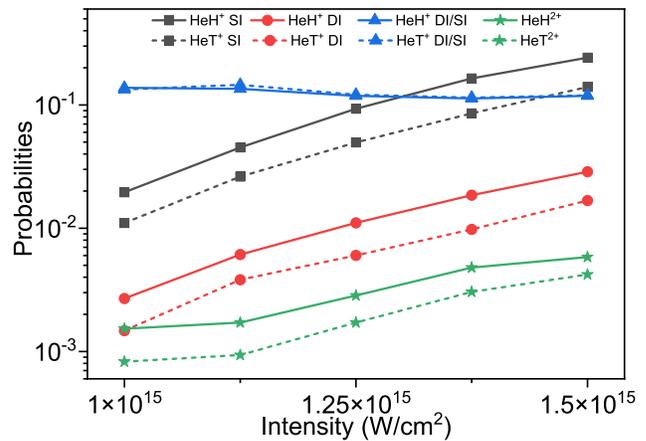}}}
\end{center}
\caption{Probabilities of SI and DI of HeH$^+$ and HeT$^+$ vs laser intensity at $\lambda=500$ nm.
Ratios of DI to SI and ionization probabilities for 1D HeH$^{2+}$ and HeT$^{2+}$ 
are also plotted here.
} \label{fig.21}
\end{figure}

\begin{figure}[t]
\begin{center}
\rotatebox{0}{\resizebox *{8.5cm}{8cm} {\includegraphics {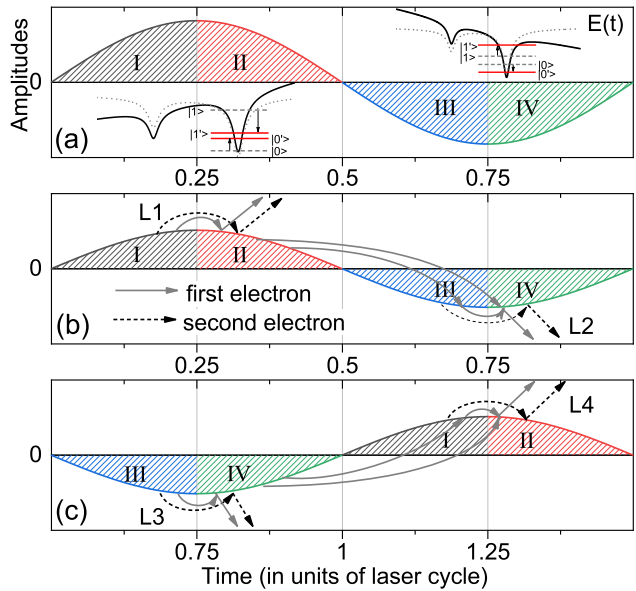}}}
\end{center}
\caption{Sketches of the laser field $E(t)$ (shaded area) in one laser cycle (a) and  motions of electrons corresponding to possible DI routes for HeH$^+$ or HeT$^+$ (b-c).
In (a), the one-cycle time region is divided into four parts (I-IV), labeled  by  different colors. The insets in (a) plot
the laser-dressed asymmetric Coulomb potential, the laser-dressed electronic states $|0'\rangle$ and $|1'\rangle$ corresponding to the free-field electronic ground state $|0\rangle$ and the first excited state $|1\rangle$ of HeH$^+$ or HeT$^+$,
when the laser polarization is antiparallel (regions I and II) or  parallel (regions III and IV) to
the permanent dipole which is directing from the He nucleus to the H(T) nucleus. For the antiparallel case, the electronic
ground state is dressed up and
the first excited state is dressed down, making the ionization easier to occur. This situation reverses for the parallel case.
As a result, the ionization is strong (weak) in the first (second) half laser cycle for the present cases. These analyses
are also applicable for HeH$^{2+}$ or HeT$^{2+}$.
In (b) and (c), possible sequential (L1 and L3) and non-sequential (L2 and L4) DI routes associated with the first electron born in regions II (b) and IV (c) are plotted.
} \label{fig.3}
\end{figure}

\begin{figure}[t]
\begin{center}
\rotatebox{0}{\resizebox *{8.5cm}{6.5cm} {\includegraphics {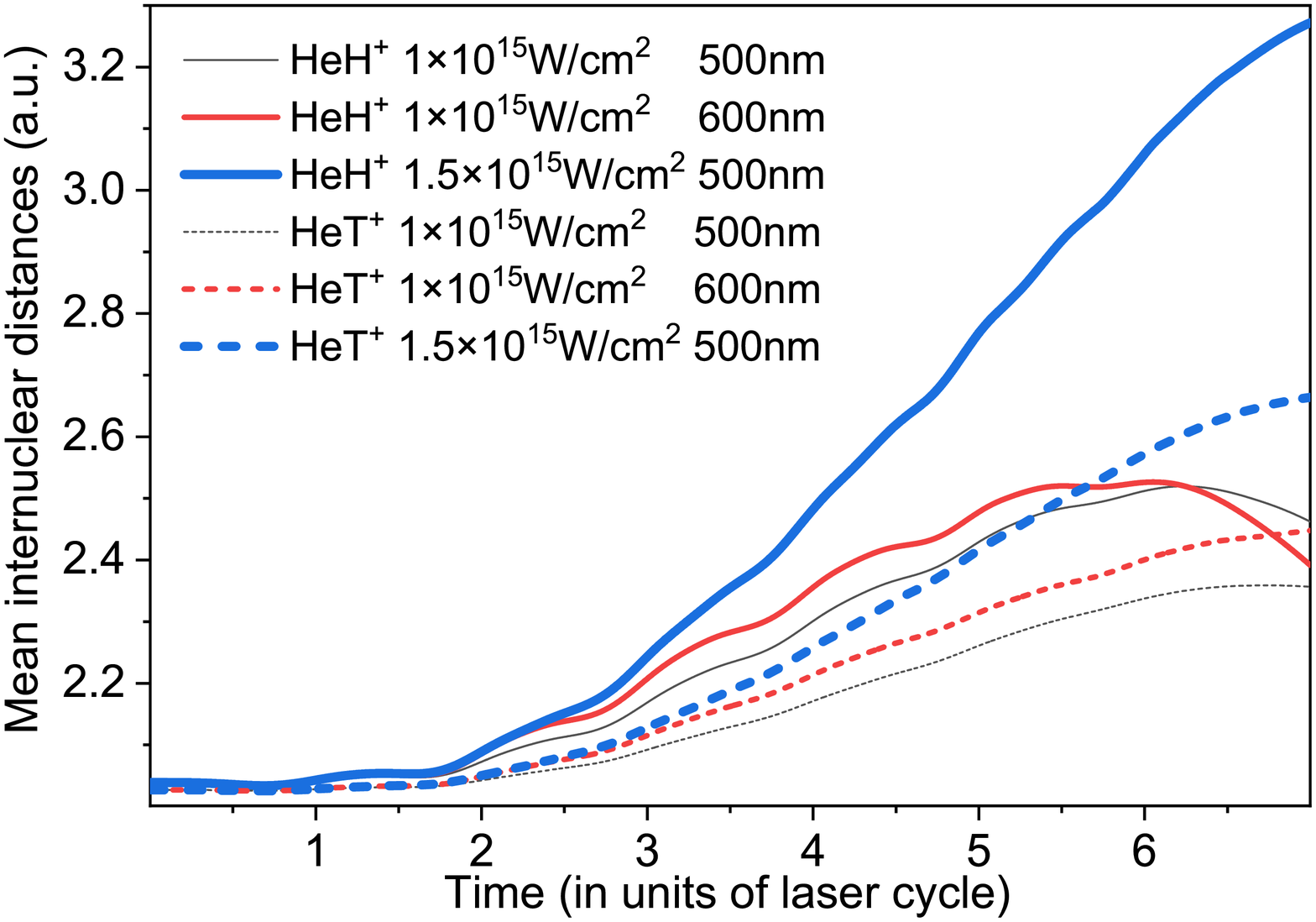}}}
\end{center}
\caption{Time-dependent mean internuclear distances of
HeH$^+$ versus HeT$^+$. The laser parameters are as in Fig. \ref{fig.1}.
} \label{fig.31}
\end{figure}

A simple evaluation on the drift momentum of the electron with the classical expression $p=-A(t)$ \cite{Corkum1993} tells that
 electrons born in  regions I and IV (II and III) have minus (plus) momenta. Here, $A(t)$ is the vector potential of $E(t)$.
The time-dependent asymmetric ionization results in Fig. 2 thus imply that for  SI, electrons with plus momenta have large amplitudes.
This point has been  shown in Ref. \cite{Wang2019} for vibrating HeH$^+$ with two-dimensional single-electron dynamics.
Here, we also show   PMDs of SI for the vibrating two-electron system of HeH$^+$ (HeT$^+$) as the insets in Fig. 2.
The insets clearly show the asymmetric PMD of SI where distributions for plus momenta show larger amplitudes.
This asymmetry in PMD of SI is more remarkable for HeT$^+$ than for HeH$^+$.

Combing the PMD results in Fig. \ref{fig.1} and Fig. \ref{fig.2}, we arrive at the conclusion that
these two electrons in DI prefer to release together along the H side when the first electron enjoys escaping along the He side in SI.
From the analyses of the ionization in Fig. \ref{fig.2}, one can also conclude that if  sequential DI (SDI) occurs,
 PMDs of DI for HeH$^+$ or HeT$^+$ will show large amplitudes in the first quadrant, similar to SI.
We therefore anticipate that non-sequential double ionization (NSDI) dominates in present cases.
In fact, extended TDSE simulations for 1D vibrating HeH$^{2+}$ (HeT$^{2+}$)
show that for the present laser parameters,
the ionization probability of HeH$^{2+}$ (HeT$^{2+}$) from its ground state
is remarkably lower than the ratio of DI to SI for HeH$^+$ (HeT$^+$), as shown in Fig. \ref{fig.21},
suggesting  that NDSI dominates here \cite{Lein2002}.
Next, we discuss possible routes of NSDI in detail.

\subsection{Mechanisms of DI}

In Fig. \ref{fig.3}(a), we plot the electric field E(t) in one laser cycle, with dividing the time region into four parts as in Fig. \ref{fig.2}.
The laser-dressed Coulomb potential and the laser-dressed two lowest electronic states of HeH$^+$ (HeT$^+$)
corresponding to the first and the second half
laser cycle are also plotted here as the insets. When the SI mainly occurs in regions II and IV, as discussed in Fig. \ref{fig.2},
we focus on possible DI routes associated with SI events in these two regions.

First, in Fig. \ref{fig.3}(b), we plot DI routes associated with the birth of the first electron in region II.
In this case,  as the first electron ionizes, the second electron can be ionized directly by the external field in this region
with contributing to the first quadrant in PMDs of DI.
It should be noted that due to the Coulomb induced large ionization time lag, these two ionized electrons in region II can find their origins
in region I. We denote these SDI routes contributing to the first quadrant ``L1".
The first electron  born in region II can also return to and recollide with the second electron in region III (short NDSI route)
and region IV (long one).
For the short route, the Coulomb effect will also induce the delay of DI time, just as it does in region I,
resulting in the emission of these two electrons in region IV and contributing to DI in the third quadrant.
For the long one, both electrons will also contribute to DI in the third quadrant. 
We denote these NSDI routes contributing to the third quadrant ``L2".
For lower laser intensities, probabilities for direct ionization of the second electron by the laser field are small,
and the route  L2
dominates in DI for the cases in Fig. \ref{fig.3}(b).

For DI routes associated with the birth of the first electron in region IV, this situation reverses, as plotted in Fig. \ref{fig.3}(c).
In this case, NSDI routes associated with the rescattering of the first electron contribute to the first quadrant and
SDI routes related to sequential ionization of these two electrons
contribute to the third quadrant.
We denote these SDI and NSDI routes ``L3" and ``L4" respectively.
Due to that the SI amplitudes are smaller in region IV than those in region II
and the second electron is bounded more deeply in region IV  than in region II (see the insets in Fig. 2(a)), for a laser cycle,
the main contributions to DI come from route L2, with  PMDs of DI showing large amplitudes in the third quadrant.
When the laser intensity increases, the ionization yields of HeH$^{2+}$ associated with direct ionization by the laser field increase
and the contributions of  route L1 increase. As a result,  PMDs of DI for the asymmetric system become more symmetric.
For increasing the  laser wavelength,  the asymmetric system stretches to a large distance at which  direct ionization is usually
easier to occur, resulting in  somewhat similar results as increasing the laser intensity. 
\begin{figure}[t]
\begin{center}
\rotatebox{0}{\resizebox *{8.5cm}{6cm} {\includegraphics {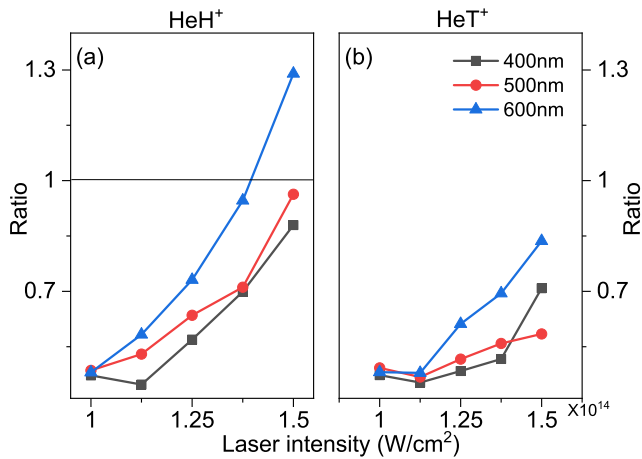}}}
\end{center}
\caption{Ratio of  PMD amplitudes of DI in the 1st quadrant to those in the 3rd quadrant for HeH$^+$ (a)
and HeT$^+$ (b) at different laser parameters as shown.
} \label{fig.4}
\end{figure}

For HeT$^+$ with heavier nuclei than HeH$^+$, at the same laser parameters, the laser-induced stretching for HeT$^+$
is smaller than for HeH$^+$, as shown in Fig. \ref{fig.31}. Generally, direct ionization prefers larger R at which the ionization potential of
the vibrating system is lower.
Accordingly,  direct-ionization yields for HeT$^+$ are also smaller than for HeH$^+$.
As a result, PMDs of DI for HeT$^+$ usually show a stronger asymmetry than for  HeH$^+$.

To validate these above discussions, in Fig. \ref{fig.4},
we plot the ratio of   amplitudes of  PMDs of DI in the first quadrant to those in the third quadrant.
One can observe from Fig. \ref{fig.4}, as  increasing the laser intensity or wavelength,  for both isotope cases, this ratio increases.
In particular,  for the same laser parameters, this ratio of  $\mathrm{HeH}^{+}$ is usually larger than that of  $\mathrm{HeT}^{+}$.
These results are in agreement with our above analyses.

\subsection{R-resolved DI and SI}

\begin{figure}[t]
\begin{center}
\rotatebox{0}{\resizebox *{8.5cm}{6.5cm} {\includegraphics {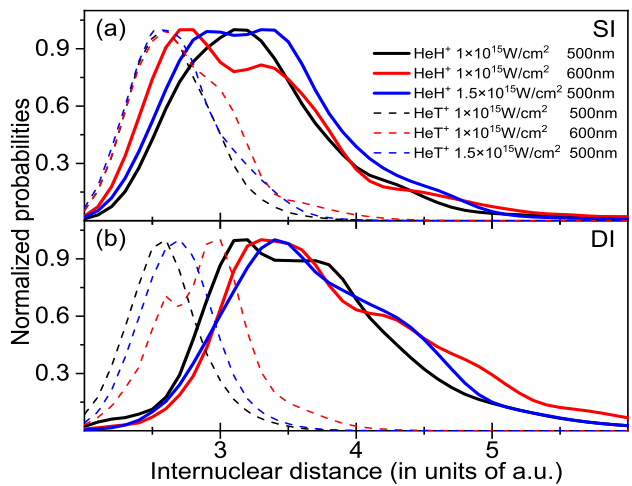}}}
\end{center}
\caption{Comparisons between R-dependent ionization probabilities $\gamma_s(R)$ of  SI (a) versus $\gamma_d(R)$ of DI (b) for HeH$^+$
and HeT$^+$, obtained with different laser parameters as shown.
} \label{fig.5}
\end{figure}

To provide further insight into electron-nucleus coupled DI dynamics of polar molecules,
in Fig. \ref{fig.5}, we plot  R-dependent probabilities  $\gamma_s(R)$ of SI and
 $\gamma_d(R)$ of DI for HeH$^+$ and HeT$^+$, averaged by the corresponding total probabilities $P_s$ and $P_d$, respectively,
at different laser parameters.

First, as increasing laser intensities and wavelengthes, for SI,
the structure  of these R-dependent distributions in Fig. \ref{fig.5} changes from a relatively sharp hump
to the plane and broad one, but the center  of the hump does not change basically.
For DI, however, the distributions extend to somewhat larger distances.
These different responses of  DI and SI on laser parameters revealed here
 agree with the experimental results in \cite{Wustelt2018}.
Secondly, in some  cases such as for HeH$^+$, the location of the hump
both for SI and DI is remarkably larger than the equilibrium separation $R_e=2$ a.u..
By comparison, for the symmetric case of model H$_2$ \cite{Lein2002},
the position of the hump is nearer to the equilibrium distance and the structure of the hump is not sensitive to laser parameters.
The results suggest that due to the permanent-dipole induced rapid stretching which
remarkably diminishes the ionization potential of the target,
both DI and SI of the asymmetric system prefer to occur at larger R when the laser intensity is not very high.
Thirdly, for the same laser parameters, the position of the hump for SI or DI of
HeH$^+$ is larger  than that for HeT$^+$, suggesting that the stretching of HeH$^+$
with lighter nuclei  is stronger than for HeT$^+$, in agreement with  previous discussions \cite{Li2019}.
Fourthly, on the whole, the position of the hump for DI of HeH$^+$ or HeT$^+$ is near to  the corresponding one for SI
at lower laser intensities and shorter wavelengthes and is somewhat larger than that at higher and longer ones.
The results suggest that at relatively high laser intensities and long  wavelengthes, SI and DI events prefer to occur
at times with a larger relative time delay.
Such events  are expected to be mainly associated with SDI processes.
For these processes, after the first electron ionizes,
the asymmetric system stretches for a while with lowering its ionization potential remarkably,
then the second electron is ionized  by the laser field, resulting in a SDI event.
These analyses also support our above discussions that for HeH$^+$ and HeT$^+$,
SDI events  increase as increasing laser intensities and wavelengthes.

\begin{figure}[t]
\begin{center}
\rotatebox{0}{\resizebox *{8.5cm}{9cm} {\includegraphics {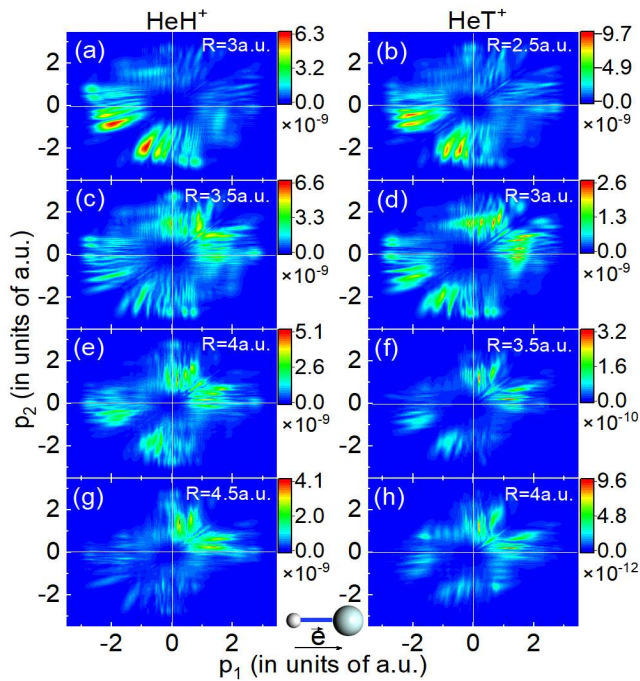}}}
\end{center}
\caption{R-dependent PMDs $\beta_d(R)$ of DI  for HeH$^+$ (the left column)
and HeT$^+$ (right).  The laser parameters are as in Figs. \ref{fig.1}(e) and \ref{fig.1}(f).
} \label{fig.8}
\end{figure}

In Fig. \ref{fig.8}, we also plot R-dependent PMDs  $\beta_d(R)$ of DI
for HeH$^+$ and HeT$^+$ at some
 distances R at which the function  $\gamma_d(R)$  has  larger amplitudes (see Fig. \ref{fig.5}).
We have chosen the laser-parameter cases in Figs. \ref{fig.1}(e) and \ref{fig.1}(f) where the distributions in the first quadrant
have relatively large amplitudes. For both cases of HeH$^+$ and HeT$^+$, one can observe from Fig. \ref{fig.8}, the main contributions
to the third (first) quadrant come from smaller (larger) R, at which the ionization potential of the system is higher (lower) and
we expect that NSDI (SDI) dominates.

\section{conclusion}
In conclusion, we have studied the ionization dynamics of vibrating HeH$^+$ with comparing it to HeT$^+$.
The photoelectron momentum distributions for both DI and SI
 show an asymmetric structure but with the contrary trend.
As the Coulomb induced large ionization time delay plays an important role in the asymmetry in SI,
the rescattering of the first electron along with this delay contributes to the asymmetry in DI.
The nuclear motion mainly influences the events of SDI. The contributions of SDI
decreases the asymmetry in DI  momentum distributions, and this decreases is more remarkable for HeH$^+$
with lighter nuclei and more rapid nuclear motion than for HeT$^+$.
This asymmetry in DI is expected to appear for other oriented polar molecules with a large permanent dipole.
In addition, the proposed DI mechanism of rescattering followed by Coulomb induced ionization time delay holds
for all of atoms and molecules including symmetric and asymmetric ones.
For these symmetric cases without a permanent dipole, this asymmetry discussed in the paper does not
appear in PMDs of DI. However, effects relating to this mechanism
are possible to resolve with using  two-dimensional  laser fields
which have shown the capability in probing sub-cycle strong-field electron dynamics.

\section*{Acknowledgement}
This work is financially supported by
the National Key Research and Development Program of China (Grant No. 2018YFB0504400);
the National Natural Science Foundation of China (Grant Nos. 91750111 and 11904072);
the Research Team of Quantum Many-body Theory and Quantum Control in Shaanxi Province, China (Grant No. 2017KCT-12);
and the Fundamental Research Funds for the Central Universities, China (Grant Nos. 2017TS008 and GK201801009).

\end{document}